\documentclass[aps,pra,twocolumn,groupedaddress,floatfix]{revtex4}%
\usepackage{graphicx}
\usepackage{dcolumn}
\usepackage{bm}
\usepackage{hyperref}
\usepackage{amssymb}
\usepackage{amsmath}
\usepackage{amsfonts}
\usepackage{amssymb}
\usepackage{color}%

\begin{document}
\title{State-to-state endothermic and nearly thermoneutral reactions in an ultracold
atom-dimer mixture}
\author{Jun Rui}
\thanks{These authors contributed equally to this work.}
\affiliation{Shanghai Branch, National Laboratory for Physical Sciences at Microscale and
Department of Modern Physics, University of Science and Technology of China,
Hefei, Anhui 230026, China}
\affiliation{CAS Center for Excellence and Synergetic Innovation Center in Quantum
Information and Quantum Physics, University of Science and Technology of
China, Shanghai 201315, China}
\affiliation{CAS-Alibaba Quantum Computing Laboratory, Shanghai 201315, China}
\author{Huan Yang}
\thanks{These authors contributed equally to this work.}
\affiliation{Shanghai Branch, National Laboratory for Physical Sciences at Microscale and
Department of Modern Physics, University of Science and Technology of China,
Hefei, Anhui 230026, China}
\affiliation{CAS Center for Excellence and Synergetic Innovation Center in Quantum
Information and Quantum Physics, University of Science and Technology of
China, Shanghai 201315, China}
\affiliation{CAS-Alibaba Quantum Computing Laboratory, Shanghai 201315, China}
\author{Lan Liu}
\affiliation{Shanghai Branch, National Laboratory for Physical Sciences at Microscale and
Department of Modern Physics, University of Science and Technology of China,
Hefei, Anhui 230026, China}
\affiliation{CAS Center for Excellence and Synergetic Innovation Center in Quantum
Information and Quantum Physics, University of Science and Technology of
China, Shanghai 201315, China}
\affiliation{CAS-Alibaba Quantum Computing Laboratory, Shanghai 201315, China}
\author{De-Chao Zhang}
\affiliation{Shanghai Branch, National Laboratory for Physical Sciences at Microscale and
Department of Modern Physics, University of Science and Technology of China,
Hefei, Anhui 230026, China}
\affiliation{CAS Center for Excellence and Synergetic Innovation Center in Quantum
Information and Quantum Physics, University of Science and Technology of
China, Shanghai 201315, China}
\affiliation{CAS-Alibaba Quantum Computing Laboratory, Shanghai 201315, China}
\author{Ya-Xiong Liu}
\affiliation{Shanghai Branch, National Laboratory for Physical Sciences at Microscale and
Department of Modern Physics, University of Science and Technology of China,
Hefei, Anhui 230026, China}
\affiliation{CAS Center for Excellence and Synergetic Innovation Center in Quantum
Information and Quantum Physics, University of Science and Technology of
China, Shanghai 201315, China}
\affiliation{CAS-Alibaba Quantum Computing Laboratory, Shanghai 201315, China}
\author{Jue Nan}
\affiliation{Shanghai Branch, National Laboratory for Physical Sciences at Microscale and
Department of Modern Physics, University of Science and Technology of China,
Hefei, Anhui 230026, China}
\affiliation{CAS Center for Excellence and Synergetic Innovation Center in Quantum
Information and Quantum Physics, University of Science and Technology of
China, Shanghai 201315, China}
\affiliation{CAS-Alibaba Quantum Computing Laboratory, Shanghai 201315, China}
\author{Bo Zhao}
\affiliation{Shanghai Branch, National Laboratory for Physical Sciences at Microscale and
Department of Modern Physics, University of Science and Technology of China,
Hefei, Anhui 230026, China}
\affiliation{CAS Center for Excellence and Synergetic Innovation Center in Quantum
Information and Quantum Physics, University of Science and Technology of
China, Shanghai 201315, China}
\affiliation{CAS-Alibaba Quantum Computing Laboratory, Shanghai 201315, China}
\author{Jian-Wei Pan}
\affiliation{Shanghai Branch, National Laboratory for Physical Sciences at Microscale and
Department of Modern Physics, University of Science and Technology of China,
Hefei, Anhui 230026, China}
\affiliation{CAS Center for Excellence and Synergetic Innovation Center in Quantum
Information and Quantum Physics, University of Science and Technology of
China, Shanghai 201315, China}
\affiliation{CAS-Alibaba Quantum Computing Laboratory, Shanghai 201315, China}

\maketitle

\textbf{Chemical reactions at ultracold temperature provide an ideal platform to study  chemical
reactivity at the fundamental level, and to understand how chemical reactions are governed by quantum mechanics \cite{Stwalley2004,Krems2008,Bell2009,Balakrishnan2016}. Recent years have witnessed the remarkable progress in studying ultracold chemistry with ultracold molecules \cite{Zahzam2006,Staanum2006,Hudson2008,Wang2013,Ospelkaus2010a,Ni2010,Miranda2011,Knoop2010,Lompe2010,Rui2017,Drews2017}. However, these works were limited to exothermic reactions. The direct observation of state-to-state ultracold endothermic reaction remains elusive. Here we report on the
investigation of endothermic and nearly thermoneutral atom-exchange
reactions in an ultracold atom-dimer mixture. By developing an indirect
reactant-preparation method based on a molecular bound-bound transition, we
are able to directly observe a universal endothermic reaction with tunable energy threshold and study the state-to-state reaction dynamics.
The reaction rate coefficients show a strikingly threshold phenomenon.
The influence of the reverse reaction on the reaction dynamics is observed for the endothermic
and nearly thermoneutral reactions. We carry out zero-range quantum mechanical scattering calculations to obtain the
reaction rate coefficients, and the three-body parameter is determined by
comparison with the experiments. The observed endothermic and nearly thermoneutral reaction
may be employed to implement collisional Sisyphus cooling of molecules \cite{Zhao2012}, study the chemical reactions in degenerate quantum gases
\cite{Carr2009,Moore2002} and conduct quantum simulation of Kondo effect
with ultracold atoms \cite{Bauer2013,Nishida2013}.}

Ultracold molecules represent great opportunities to explore cold controlled chemistry and study state-to-state reaction dynamics at the quantum level \cite{Krems2008,Carr2009,Quemener2012}. Recently, with the
development of ultracold molecule techniques, significant progress has
been achieved in studying ultracold bimolecular reactions.  Molecule-molecule \cite{Ospelkaus2010a,Ni2010,Miranda2011} and
atom-dimer \cite{Knoop2010,Lompe2010,Rui2017} atom-exchange reactions have been observed
at temperatures below 1 $\mu$K.  However, these
studies are limited to exothermic reactions, and ultracold reactions in the
endothermic regime remain unexplored. Endothermic reactions are qualitatively
different from exothermic reactions due to the presence of energy thresholds.
Therefore ultracold endothermic reactions are expected to show
threshold phenomenon, i.e., the reaction rate coefficients decrease remarkably when the threshold is changed from zero to a small positive value. Further, for ultracold endothermic reactions, the
kinetic energy converts into internal energy and thus the resultant
products are colder than reactants, which means that both the reactants and
products should be described by quantum mechanics. This is in contrast to most
of ultracold exothermic reactions where only the reactants are cold, whereas
the products are much hotter due to the large exothermicity, and consequently
the products can still be described by classical treatment \cite{Herschbach2009}.

Exothermic reactions in ultracold gases always lead to detrimental effects, such as loss of the molecules or heating of the samples, and
thus they are expected to be suppressed~\cite{Balakrishnan2016}. In contrast, endothermic
reactions do not cause heating and may even result in cold products. It has been proposed that endothermic inelastic collision may be employed to implement collisional Sisyphus cooling of molecules~\cite{Zhao2012}. Moreover,
endothermic reactions may offer a platform to study superchemistry where
quantum degeneracy may strongly affect the reaction \cite{Moore2002,Carr2009}, since they will not destroy quantum degeneracy due to heating.
Besides, the thermoneutral
reaction (zero energy change, neither exothermic nor endothermic) may be employed to implement quantum simulation of the Kondo effect with
ultracold atoms, as proposed in reference \cite{Bauer2013}.

Exploring endothermic reactions at nearly zero temperature are challenging,
because the collision energies of the reactants are much smaller than the
threshold of most of endothermic reactions and thus the reaction should be
simply forbidden. Theoretical schemes have been proposed to use laser to
control an isotope-exchange reaction to be endothermic \cite{Tomza2015}. These
techniques remain to be demonstrated. So far, the universal atom-exchange reactions
with weakly bound Feshbach molecules \cite{Knoop2010,Lompe2010,Rui2017}
offer a unique possibility to study the endothermic reaction at nearly zero
temperature. The reaction can be tuned to be exothermic, endothermic,
or thermoneutral by simply varying the magnetic field.

Despite the easy tunability of the energetics, the direct observation of the state-to-state universal
endothermic reactions and measurement of the reaction rate remain unsolved
questions. In Refs. \cite{Knoop2010,Lompe2010}, only the overall loss rates of the molecule
reactants are observed. Although in Ref. \cite{Knoop2010} it is claimed that a pronounced threshold behaviour is observed in the endothermic
regime, the observed overall loss rate includes the contribution from both the endothermic exchange reaction and the exothermic vibrational
relaxation into deeply bound states, and the theoretical model has shown that dominant loss mechanism is actually the exothermic vibrational
relaxation. In Ref. \cite{Rui2017}, the state-to-state reaction rate can only
be measured in the exothermic regime with large released energy.

\begin{figure}[ptb]
\centering
\includegraphics[width=8cm]{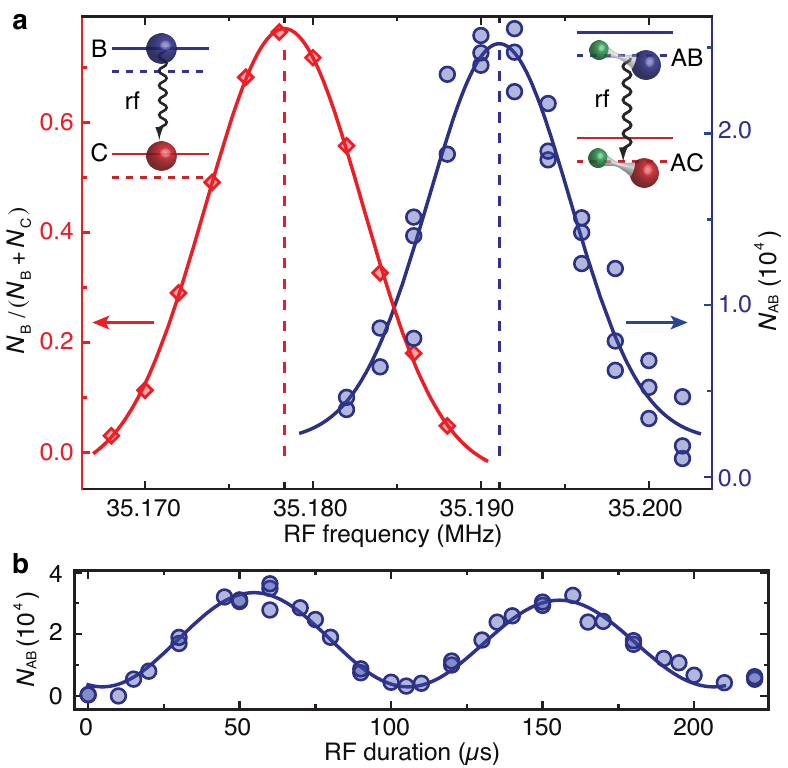}\caption{\textbf{Molecular bound-bound rf spectrum and Rabi oscillation.} \textbf{a,} The measured atomic transition spectrum (red diamonds) and molecular bound-bound spectrum (blue circles) at 130.29 G. For the atomic spectrum, the fraction of the B atoms is recorded. For the molecular spectrum, the number of molecules transferred into the AB state is recorded by imaging the dissociated K atoms in the $|9/2,-7/2\rangle$ state. The data points are fitted with Gaussian functions. \textbf{b,} The measured bound-bound Rabi oscillation between the AB and AC molecule states at 130.26 G. The bound-bound Rabi frequency is fitted to be $2\pi\times9.9(1)$ kHz.}
\label{fig1}%
\end{figure}

Here we report on the study of the state-to-state endothermic and nearly
thermoneutral reaction in an ultracold atom-dimer mixture. Our experiment starts with approximately $3.0\times10^{5}$ $^{23}$Na atoms and
$1.6\times10^{5}$ $^{40}$K atoms at about $600$ nK in a crossed-beam optical
dipole trap. The measured trap frequencies for K are $h\times(250,237,79)$ Hz,
with $h$ being the Planck constant. The Na atoms are prepared in the lowest
hyperfine $|F,m_{F}\rangle_{\text{Na}}=|1,1\rangle$ state, and the K atoms can
be prepared in any hyperfine ground state $|F,m_{F}\rangle_{\text{K}%
}=|9/2,-9/2\rangle...|9/2,9/2\rangle$.

At a magnetic field near 130 G, the
interspecies Feshbach resonance between $|1,1\rangle$ and $|9/2,-3/2\rangle$
and the Feshbach resonance between $|1,1\rangle$ and $|9/2,-5/2\rangle$ overlap (Supplementary Information). The overlapping
Feshbach resonances allow studying the universal atom-exchange reaction between weakly
bound Feshbach molecules and atoms \cite{Incao2009,Knoop2010,Lompe2010,Rui2017}. In our experiment, the reaction may be described by $\mathrm{{AB+C\rightarrow AC+B,}}$ where A
denotes $^{23}$Na atom in the $|1,1\rangle$ state, and B and C denote $^{40}$K
atoms in the $|9/2,-5/2\rangle$ and $|9/2,-3/2\rangle$ state respectively. The
AB and AC molecules are the corresponding NaK Feshbach molecules with binding
energies $E_{\mathrm{{AB}}}^{\mathrm{{b}}}$ and $E_{\mathrm{{AC}}%
}^{\mathrm{{b}}}$, respectively. The energy change in the reaction is given by $\Delta E=E_{\mathrm{{AB}}}^{\mathrm{{b}}}-E_{\mathrm{{AC}}}^{\mathrm{{b}}}$. The state-to-state reaction dynamics can be measured by
preparing the reactant mixture and then monitoring the time evolution of AB
reactant and B product, and the reaction rate coefficient is extracted from the
time evolution. For detection, the AB
molecules are dissociated into the $\mathrm{{A}+|9/2,-7/2\rangle}$ state and
the B atoms are transferred to the $|9/2,-7/2\rangle$ state by rf pulses, and
the AB molecule number and B atom number are determined by measuring the atom number in the $|9/2,-7/2\rangle$ state, respectively.

\begin{figure}[ptb]
\centering
\includegraphics[width=8cm]{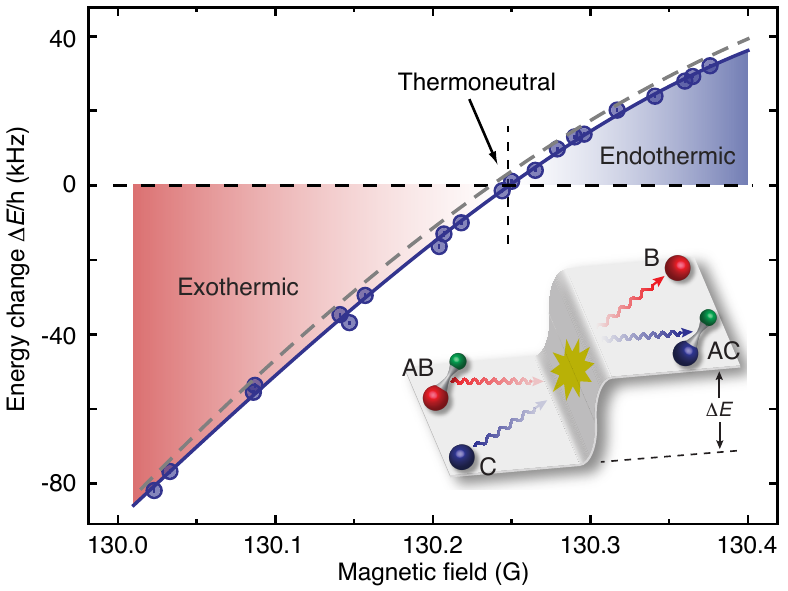}\caption{\textbf{The reaction energy changes versus magnetic field.} The measured
energy change $\Delta E$ for the reaction $\mathrm{AB+C \rightarrow AC+B}$ as
a function of the magnetic field. The energy change is precisely determined by
measuring the molecular bound-bound transition frequency and the free
atom-atom transition frequency using rf spectroscopy. The blue solid line is a
polynomial fit to the data points. The vertical dashed line marks the magnetic
field $B_{\mathrm{{tn}}}=130.249(2)$ G at which the reaction is thermoneutral
with zero energy change. For $B<B_{\mathrm{{tn}}}$ the reaction is exothermic,
and for $B>B_{\mathrm{{tn}}}$, the reaction is endothermic. As a comparison,
the grey dashed line shows the energy change obtained from the difference
between the two binding energy curves. Error bars inside the circles represent
$\pm1$ s.d.}%
\label{fig2}%
\end{figure}

In all previous works studying ultracold reaction with Feshbach molecules, the
molecule reactants are directly formed using magnetic field association \cite{Knoop2010,Lompe2010} or
radio frequency association \cite{Rui2017}. However, these methods cannot measure the
reaction rate in the endothermic regime, either due to the low formation
efficiency or large background noise caused by the preparation process.

In our system, the primary experimental challenge is to efficiently prepare
the reactants $\mathrm{{AB+C}}$ at the magnetic fields where endothermic
reactions with small $\Delta E$ may occur.  For the exothermic reactions with large and negative $\Delta
E,$ the reactants can be prepared by radio frequency (rf) association
\cite{Wu2012,Rui2017} of the AB molecule from an $\mathrm{{A+C}}$ mixture.
However, this preparation method is not applicable for small $|\Delta E|$,
because in this regime the scattering lengths $a_{\mathrm{{AB}}}$ and
$a_{\mathrm{{AC}}}$ in the two scattering channels are close to each other.
This results in a rather low association efficiency since the bound-free
Franck-Condon factor $F_{\mathrm{{bf}}}\varpropto(1-a_{\mathrm{{AC}}%
}/a_{\mathrm{{AB}}})^{2}$ is largely suppressed \cite{Chin2005}.

To circumvent this problem, we develop an indirect reactant-preparation
method based on the molecular bound-bound transition. Instead of directly
forming the AB molecules, we first associate the AC molecules from the
$\mathrm{{A}+|9/2,-1/2\rangle}$ mixture and then transfer the weakly bound
molecules from the AC to the AB state by an rf $\pi$ pulse. A subsequent rf
$\pi$ pulse transferring the free K atoms from the $|9/2,-1/2\rangle$ state to
the C state then prepares the desired reactants.  The idea behind this method is that for a small $|\Delta
E|$, the suppression of bound-free Franck-Condon factor is accompanied by an
enhancement of the bound-bound Franck-Condon factor \cite{Chin2005} $F_{\mathrm{{bb}}%
}\varpropto$ $4a_{\mathrm{{AC}}}a_{\mathrm{{AB}}}/(a_{\mathrm{{AC}}%
}+a_{\mathrm{{AB}}})^{2}$. Therefore, although it is difficult
to form the AB molecule from the $\mathrm{{A+C}}$ scattering state, it does
allow us to create the AB molecule from the AC bound state.

The Feshbach resonance between the A and C atoms tends to be closed-channel dominated,
and thus the AC molecules can only be efficiently associated when close to the
resonance. In our experiment, we can efficiently form the AC molecules in a
small magnetic field window between $130.02$ G and $130.38$ G using Raman photoassociation (Supplementary Information). The bound-bound rf spectrum and Rabi oscillation are measured, as shown in Fig. \ref{fig1}. The difference
between the bound-bound transition frequency and the free-free transition frequency precisely determines the energy change $\Delta
E=h(\nu_{\mathrm{{AC}\rightarrow{AB}}}-\nu_{\mathrm{{C}\rightarrow{B}}})$ of
the reaction, which is in the range $h\times(-82,32)$ kHz. The measured
$\Delta E$ as a function of the magnetic field is shown in Fig. \ref{fig2}.
The polynomial fit yields $\Delta E=0$ at $B_{\mathrm{{tn}}}=130.249(2)$ G,
which agrees well with the value of $130.24(1)$ G determined from the
intersection point of the binding energy curves.  The bound-bound transition has an efficiency of better than 95\%, thanks to the
enhanced bound-bound Franck-Condon factor, and approximately
$2\times10^{4}$ AB molecules are formed. After preparing the reactants, the
time evolutions of the AB molecule reactants and the B atom products are recorded.

\begin{figure*}[pth]
\centering
\includegraphics[width=16cm]{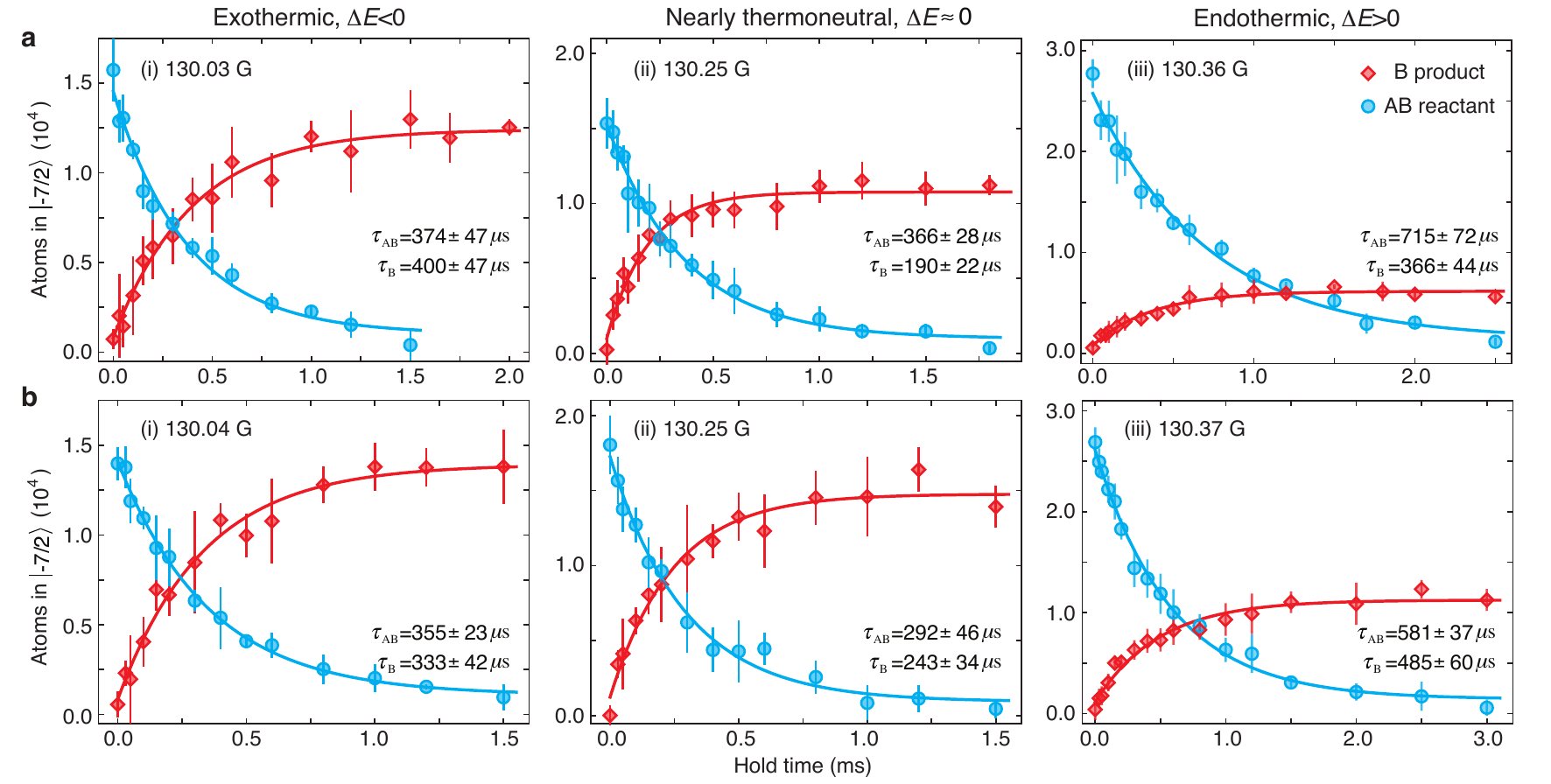}\caption{\textbf{The reaction dynamics.} \textbf{a,} The decay of the AB reactant
(blue circles) and the increase of the B product (red diamonds) as a function
of the hold time, for (i) the exothermic reaction at $B=130.03$ G with $\Delta
E=-76.7(3)$ kHz, (ii) the nearly thermoneutral reaction at $B=130.25$ G with
$\Delta E=1.1(2)$ kHz, and (iii) the endothermic reaction at $B=130.36$ G with
$\Delta E=28.0(2)$ kHz. The blue and red solid lines are exponential fitting
curves with $1/e$ time constants $\tau_{\mathrm{{AB}}}$ and $\tau
_{\mathrm{{B}}}$, respectively. For the exothermic reaction, $\tau
_{\mathrm{{AB}}}$ and $\tau_{\mathrm{{B}}}$ agree with each other within
mutual uncertainties, whereas for the endothermic and nearly thermoneutral
reactions, $\tau_{\mathrm{{AB}}}$ and $\tau_{\mathrm{{B}}}$ are not
consistent. \textbf{b,} The time evolution of the AB reactant (blue circles)
and the B product (red diamonds) after continuously removing the AC product in
the reaction process. The two time constants $\tau_{\mathrm{{AB}}}$ and
$\tau_{\mathrm{{B}}}$ are consistent with each other within the mutual
uncertainties also for the endothermic and nearly thermoneutral reactions, and
the increased numbers of the B products are larger. Error bars represent
$\pm1$ s.d.}%
\label{fig3}%
\end{figure*}

The reaction dynamics are shown in Fig.~\ref{fig3}. For magnetic fields larger than $B_{\mathrm{{tn}}}$, the reaction is
endothermic with $\Delta E>0$. With increasing magnetic field the reaction
threshold increases and the maximum number of the B products decreases
quickly. Using the indirect preparation method, the endothermic reaction is
clearly observed up to $130.38$ G with a reaction threshold of $32$ kHz.

At $130.25$ G, the observed reaction is a nearly thermoneutral reaction with
the measured energy change $\Delta E=h\times1.1(2)$ kHz. The ideal
thermoneutral reaction has zero energy change \cite{Upadhyay2006}, i.e.,
$\Delta E=0$. Previously, the thermoneutral reaction usually refers to the
\textquotedblleft reaction" that the reactants and products are identical,
e.g., $\mathrm{{H_{2}+H\rightarrow H_{2}+H}}$, where $\mathrm{{H_{2}}}$ and H
are the hydrogen molecule and atom in their lowest internal states,
respectively \cite{Stwalley2004}. However, this is not an observable reaction.
In our system, by preparing the reactants at the magnetic field where the
Feshbach molecules have exactly the same binding energy, an observable
thermoneutral reaction can be realized. In the current experiment, the
achievable minimal energy change is limited by the resolution and uncertainty
of the magnetic field.

We analyze the data by fitting the time evolution of the AB reactant and B
product numbers using exponentials. As shown in Fig. \ref{fig3}a, we find that
for the exothermic reactions, the two $1/e$ time constants $\tau
_{\mathrm{{AB}}}$ and $\tau_{\mathrm{{B}}}$ agree with each other within
mutual uncertainties, whereas for the endothermic reactions and nearly
thermoneutral reactions, the two time constants are different from each other.
We attribute this discrepancy to the influence of the reverse reaction
$\mathrm{{AC+B\rightarrow AB+C}}$, i.e., the reaction is reversible.

To demonstrate that this discrepancy is caused by the reverse reaction, we
continuously remove the AC molecule products from the reaction mixture by
dissociating them into the $\mathrm{{A}+|9/2,-1/2\rangle}$ state. To do so, we
first remove the A atoms after the AB molecules have been prepared. Then during the reaction process, we use the
Raman light to continuously dissociate the AC molecules into the
$\mathrm{{A}+|9/2,-1/2\rangle}$ state (see Supplementary Information). This
procedure suppresses the reverse reaction and makes the
reaction irreversible. In this way, we find $\tau_{\mathrm{{AB}}}$ and
$\tau_{\mathrm{{B}}}$  agree with each other within mutual uncertainties
also for the endothermic and nearly thermoneutral reactions, as shown in Fig.
\ref{fig3}b.

\begin{figure}[tbh]
\centering
\includegraphics[width=8cm]{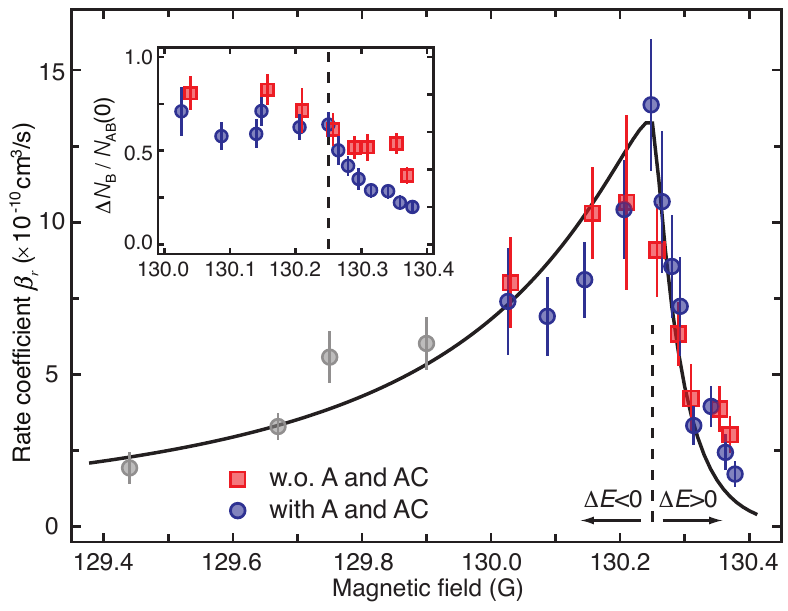}\caption{\textbf{The behaviour of the
reaction rate coefficients as a function of the magnetic field.} The blue circles and red squares
represent the data points measured by using the indirect reactant-preparation
method. The red squares are the data points obtained by continuously removing
the AC molecules (after eliminating the A atoms) when the reverse reactions
are significantly suppressed. The grey circles represent the data points
obtained from Ref. \cite{Rui2017}. The solid line is the numerical result with
the fitted three-body parameter $\Lambda=22.6/a_{0}$ and the normalization
coefficient $C_{\beta}=1.4$. The inset shows the ratios between the net
increase of the B atom product and the initial AB molecule number, where the
square points give the reaction branching ratio. Error bars represent $\pm1$
s.d.}%
\label{fig4}%
\end{figure}

We now estimate the reaction rate coefficient $\beta_{r}$ from the measured
time evolutions. Taking the reverse reaction into account, the time evolution
of the number of B products may be described by
\begin{equation}
\dot{N}_{\mathrm{{B}}}=\beta_{r}\bar{n}_{\mathrm{{C}}}N_{\mathrm{{AB}}}%
-k_{re}N_{\mathrm{{AC}}}N_{\mathrm{{B}}},\label{evo}%
\end{equation}
where the first and second terms on the right side of Eq. (\ref{evo}) account
for the forward and reverse reactions, respectively. By continuously removing
the AC products, the second term can be safely neglected. The mean density of
the C atoms is approximated by $\bar{n}_{\mathrm{{C}}}=\alpha N_{\mathrm{{C}}%
}$, with $\alpha=((m_{\mathrm{K}}\bar{\omega}^{2})/(4\pi k_{B}T))^{3/2},$
where $\bar{\omega}$ and $T$ are the geometric mean of the trap frequencies
and the temperature, respectively. We have assumed $\bar{n}_{\mathrm{{C}}}$ to
be a constant as the number of the AB molecules is about one order of
magnitude smaller than the number of the C atoms. The reverse reaction
complicates the reaction dynamics, but the reaction rate coefficient may still
be extracted using the simple equation%
\begin{equation}
\beta_{r}=\frac{\dot{N}_{\mathrm{{B}}}(0)}{\bar{n}_{\mathrm{{C}}%
}N_{\mathrm{{AB}}}(0)},\label{ratecoeff}%
\end{equation}
where $N_{\mathrm{{AB}}}(0)$ is the initial number of the AB molecules and
$\dot{N}_{\mathrm{{B}}}(0)$ is the initial derivative of the time evolution of
the B atoms. Eq. (\ref{ratecoeff}) is also applicable to the reversible
reaction, since we start from the $\mathrm{{AB+C}}$ mixture and thus at $t=0$
the products $N_{\mathrm{{AC}}}$ and $N_{\mathrm{{B}}}$ are both negligible.
We assume that $N_{\mathrm{{AB}}}(t)$ and $N_{\mathrm{{B}}}(t)$ can be
approximated by exponentials, and obtain $\dot{N}_{\mathrm{{B}}}(0)$ and
$N_{\mathrm{{AB}}}(0)$ from exponential fits of the measured data points. As
shown in Fig. \ref{fig4}, the reaction rate coefficients obtained in this way
are similar to the results in the case where the reverse reaction has been
suppressed.

For the exothermic reactions, the reaction rate coefficient
increases as the magnetic field increases. This may be qualitatively
understood from the increase of the overlap between the wave functions of the
two weakly bound molecules, when the difference between the two scattering
lengths decreases \cite{Knoop2010,Incao2009}. For the endothermic reaction,
the reaction rate coefficient shows a strikingly threshold behaviour, as expected. Besides,
the ratio between the increased number of the B products and the initial
number of the AB reactants is given in the inset of Fig. \ref{fig4}. The data
points after removing the A atoms and the AC molecules directly give the
branching ratio of the reaction.

To understand the reaction rate, we perform quantum mechanical reactive
scattering calculations using the zero-range approximation. This is justified
since in our experiments the scattering lengths $a_{\mathrm{{AB}}}$ and
$a_{\mathrm{{AC}}}$ are both much larger than the van der Waals lengths, and
the binding energies of the AB and AC molecules are well described by the
universal model (Supplementary Information). In this case, the $s$-wave
atom-dimer scattering may be described by the generalized
Skorniakov--Ter-Martirosian (STM) equations
\cite{Braaten2006,Braaten2010,Helfrich2010}. In the STM equations, the only input parameters
are the scattering lengths $a_{\mathrm{{AB}}}$ and $a_{\mathrm{{AC}}}$
(Supplementary Information), and the unknown three-body parameter $\Lambda$,
which will be determined by comparing the reaction rates.

By numerically solving the STM equations, we calculate the reaction rate
coefficients with $\Lambda$ being a fitting parameter. The results are
compared with the experiment for the magnetic fields between $129.44$ G and
$130.38$\ G, where the data points for exothermic reactions between $129.44$ G and $129.91$ G are
obtained from our previous work \cite{Rui2017}. We find that for a range of
values for $\Lambda$, the shapes of the theoretical curves and the
experimental data points qualitatively agree with each other, whereas the
measured rate coefficients are larger than the calculated ones. We attribute
this discrepancy to the systematic shifts in our data due to uncertainties in
the particle numbers, density calibration and so on. Therefore, we include a
normalization factor $C_{\beta}$ as the second fitting parameter. As shown in
Fig. \ref{fig4}, we find good agreement between the theory and the experiment
with the fitting parameters $\Lambda=22.6/a_{0}$ and $C_{\beta}=1.4$, where $a_{0}$ is the Bohr radius.

In conclusion, we have experimentally studied state-to-state endothermic and nearly
thermoneutral reactions in an ultracold atom-dimer mixture. The universal character of the reaction
allows us to understand the reaction from a zero-range
quantum mechanical scattering calculation.  The study of endothermic reaction opens up the realistic avenue to implement collisional Sisyphus cooling of molecules \cite{Zhao2012}.
The observation of thermoneutral reaction is crucial to the quantum
simulation of the Kondo effect with ultracold molecules
\cite{Bauer2013,Nishida2013}. This scheme requires the Feshbach molecules to have almost
the same binding energy \cite{Bauer2013}, which is exactly the nearly
thermoneutral regime demonstrated in our work.

We would like to thank Peng Zhang, Cheng Chin, and Matthias Weidem\"{u}ller for helpful discussions, and Hui Wang, Xing Ding
and Xiao-Tian Xu for their important assistances. We acknowledge Ingo Nosske for
carefully reading the manuscript. This work was supported by the National
Natural Science Foundation of China (under Grant No.11521063, 11274292), the
National Fundamental Research Program (under Grant No. 2013CB336800), and the
Chinese Academy of Sciences.

~\\

\noindent\rule[0.5\baselineskip]{9cm}{0.5pt}

\section*{Supplementary Information}

\subsection{Feshbach resonance}

\begin{figure}[ptb]\centering
\includegraphics[width=8cm]{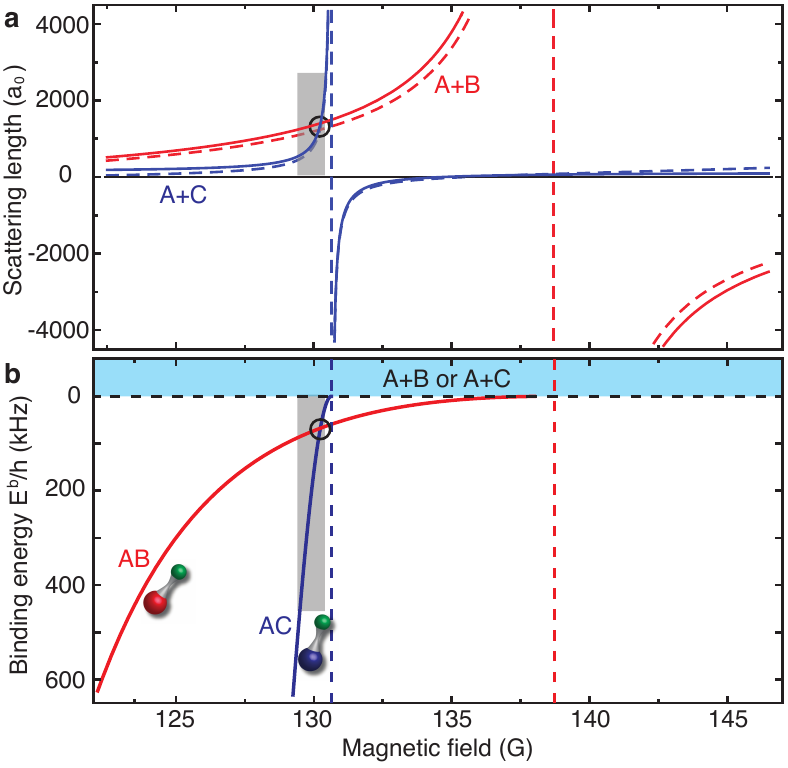}
\renewcommand\thefigure{S1}
\caption{\textbf{Overlapping Feshbach resonance.} \textbf{a}.  The scattering lengths of the collision channels $\rm A+B$ (red) and $\rm A+C$ (blue). Solid lines represent the scattering lengths from the universal model. Dashed lines represent the results from the coupled-channel calculations. \textbf{b}. The binding energies of the weakly bound AB and AC Feshbach molecules obtained from the universal model, which predict the two curves to intersect at 130.24 G. The grey shaded region denotes the range of magnetic fields in which the current experiment is performed. }\label{figs1}%
\end{figure}

The scattering length near the Feshbach resonance $\widetilde{a}%
(B)=a_{\mathrm{bg}}[1-\Delta B/(B-B_{0})]$ has been determined in Ref.
\cite{Rui2017s} by fitting the binding energies of the Feshbach molecules using
the universal model. For the Feshbach resonance between A and B, we obtain
$a_{\mathrm{bg}}=-455$ $a_{0}$, $B_{0}=138.71$ G and $\Delta B=-34.60$ G. For
the Feshbach resonance between A and C, we have $a_{\mathrm{bg}}=126$ $a_{0}$,
$B_{0}=130.637$ G and $\Delta B=4.0$ G. The binding energies of the Feshbach
molecules are given by$\ E_{\mathrm{ij}}^{\mathrm{{b}}}=\hbar^{2}/2\mu
_{d}a_{\mathrm{ij}}{}^{2}$, where $a_{\mathrm{ij}}=\widetilde{a}_{\mathrm{ij}%
}(B)-\bar{a},$ with $\bar{a}=51$ $a_{0}$ being the mean scattering length. The
scattering lengths and binding energies are shown in Fig. \ref{figs1}. Here
$a_{\mathrm{AB}}$ and $a_{\mathrm{AC}}$ are the input parameters in the STM
equations. From the binding energy curves, we obtain $E_{\mathrm{AB}%
}^{\mathrm{{b}}}=E_{\mathrm{AC}}^{\mathrm{{b}}}$ at $B=130.24(1)$ G, which is
about 10 mG smaller than $B_{tn}=130.249(2)$ G which was determined from the direct
measurement. Therefore, in the numerical calculations, the magnetic field is
shifted by 10 mG to compare with the experiments.

\subsection{Raman photoassociation and molecular bound-bound transition}

\begin{figure}[ptb]\centering
\includegraphics[width=6cm]{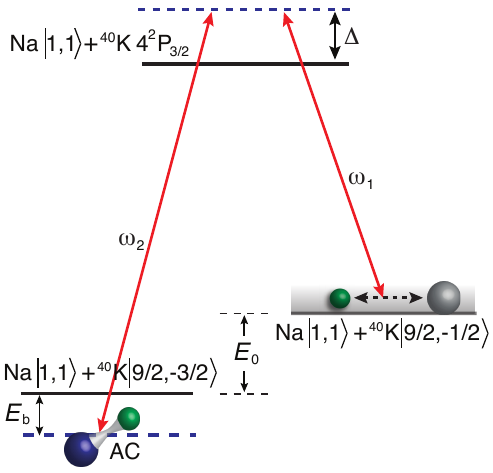} \renewcommand\thefigure{S2}
\caption{\textbf{Raman photoassociation.} The atom mixture is initially prepared in the $\rm A + |9/2,-1/2\rangle$ state. The AC molecule is formed by the Raman light. $E_b$ denotes the binding energy of the AC molecules. $E_0$ denotes the free atomic transition frequency between the two spin states of K atoms. }\label{figs2}%
\end{figure}

In this work, we employ Raman photoassociation to form AC Feshbach molecules~\cite{Fu2014s}. We use two blue-detuned Raman light fields to couple the $|9/2,-1/2\rangle$
and C states with a single-photon detuning of $\Delta\approx2\pi\times251$
GHz, with respect to the D2 line transition of the $^{40}$K atom, as shown in
Fig. \ref{figs2}. The two Raman beams have a frequency difference of
approximately $2\pi\times37$ MHz at the magnetic fields of around 130 G. Thus,
they can be simply created by using two acousto-optic modulators with the light
field from a single laser. The width of each Raman beam is about 0.8 mm at the
position of the atoms, and the power is 30 mW. The Raman Rabi frequency for
the $|9/2,-1/2\rangle\rightarrow\mathrm{{C}}$ transition is about $2\pi
\times90$ kHz.

To associate the AC molecules from the $\mathrm{A+|9/2,-1/2\rangle}$ mixture,
we use a 300-$\mu$s Gaussian pulse with a full width at half maximum (FWHM)
duration of about 150 $\mu$s. Using a Gaussian pulse, the high Fourier side
lobes can be efficiently suppressed, and thus the transfer of free atoms from
the $|9/2,-1/2\rangle$ to the C state can be neglected. The Raman light fields
have no influence on the active stabilization loop of the magnetic field, and
thus it does not cause a shift of the magnetic field.

After creating the AC molecules, we determine the molecular bound-bound
transition frequency $\mathrm{\nu_{AC\rightarrow AB}}$ by measuring the
molecule radio-frequency (rf) spectrum. This is achieved by first transferring
the AC molecule to the AB molecule via a rf pulse, and then the AB molecules
are detected by dissociating them into the $\mathrm{A+|9/2,-7/2\rangle}$ state for
imaging. Besides, we also measure the free atomic transition frequency
$\mathrm{\nu_{C\rightarrow B}}$ by measuring the rf spectrum between the C and
B states. From these measurements, the reaction
energy change $\Delta E=\nu_{\mathrm{AC\rightarrow AB}}-\nu
_{\mathrm{C\rightarrow B}}$ can be accurately determined, and the results are
shown in Fig. 1 in the main text. After determining the bound-bound
transition frequency, the Rabi oscillation between the AB and AC molecule
states is measured, and thus the duration of
the $\pi$ pulse is determined.


\subsection{Remove the remaining A atoms and the AC molecule products}

\begin{figure}[ptb]\centering
\includegraphics[width=8cm]{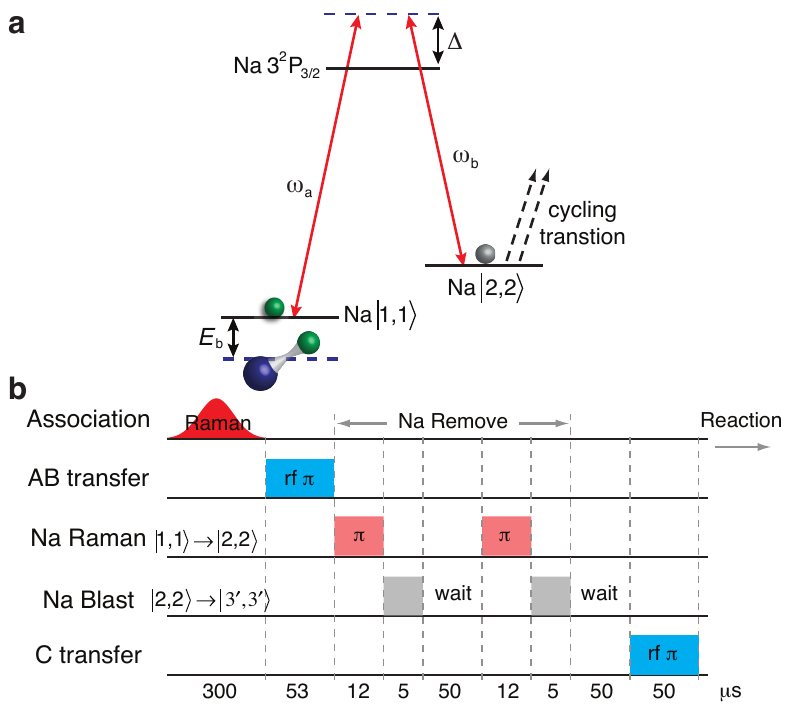} \renewcommand\thefigure{S3}
\caption{\textbf{Removal of the remaining A atoms.} \textbf{a}. The energy levels of the two-photon Raman transition used for transferring the Na atoms from the A to the $|2,2\rangle$ state. After the Raman transfer, the Na atoms in the $|2,2\rangle$ state are resonantly blasted out of the optical trap by a light pulse  resonant with the cycling transition. \textbf{b}. The time sequence of the preparation process including the removal of the remaining A atoms.}\label{figs3}%
\end{figure}

To demonstrate that the discrepancy between the time constants $\tau
_{\mathrm{{AB}}}$ and $\tau_{\mathrm{{B}}}$ for the endothermic reaction and
nearly thermoneutral reaction is caused by the reverse reaction, we remove the
AC molecule products continuously during the reaction process by dissociating
them into the $\mathrm{A+|9/2,-1/2\rangle}$ state. However, the remaining A
atoms and the residual $\mathrm{|9/2,-1/2\rangle}$ atoms due to the
imperfection of the $\pi$ pulse may cause a competition between the
dissociation and association process. Therefore, we first remove the A atoms
after the AB molecules have been prepared. This is achieved as follows. We
first transfer Na atoms from the A to the $|2,2\rangle$ state using a blue-detuned
Raman $\pi$ pulse with a single-photon detuning of $\Delta\approx2\pi
\times215$ GHz with respect to the D2 line of Na atom, and then blast them out
of the optical trap by applying a light pulse resonantly coupling the
$|2,2\rangle\rightarrow$ $|3^{\prime},3^{\prime}\rangle$ cycling transition,
as shown in Fig. \ref{figs3}a. The Raman fields for Na atoms have a width of
about 1.1 mm at the position of the atoms, and each beam has a power of about
50 mW, which yields a Raman Rabi frequency of $2\pi\times46$ kHz for the
$\mathrm{A\rightarrow|2,2\rangle}$ transition. The Raman-Blast sequences are
applied twice to completely remove the remaining A atoms. The time sequence
including  the removal of the A atoms is shown in Fig. \ref{figs3}b. The Raman-Blast
process also causes small loss of the AB molecules. Typically, we still keep
$90\%$ of the AB molecules after removing the remaining A atoms.

\begin{figure*}[ptb]\centering
\includegraphics[width=12cm]{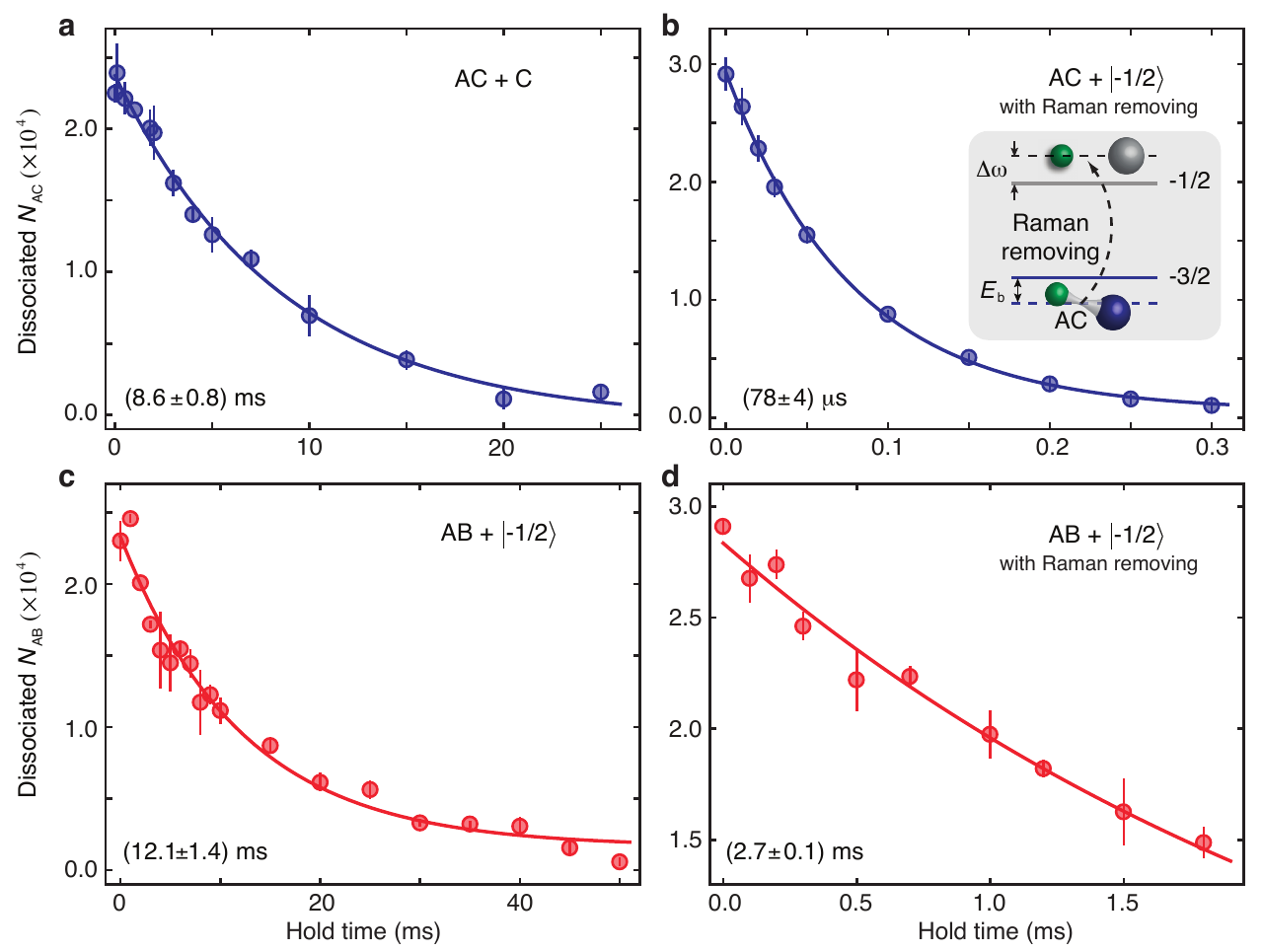} \renewcommand\thefigure{S4}
\caption{\textbf{Lifetime of the molecules under different conditions}. The measurements are taken after removing the remaining A atoms at 130.26 G. \textbf{a} and \textbf{b}. The lifetime of the AC molecules without and with the Raman removing fields, respectively. The two-photon frequency is tuned to be about 220 kHz above the bound-free transition frequency. The AC molecules are detected by first transferring them to the AB state and then dissociating them into the $\rm A+|9/2,-7/2\rangle$ state. \textbf{c} and \textbf{d}. The lifetime of the AB molecules without and with the Raman removing fields, respectively. Error bars represent $\pm1$ s.d.}\label{figs4}%
\end{figure*}

Then we can remove the AC molecule products during the reaction process by
continuously dissociating them into the $\mathrm{A+|9/2,-1/2\rangle}$ state.
This is achieved by using the Raman light coupling the
$\mathrm{|9/2,-1/2\rangle}$ and the C states, which is also used to associate the AC
molecules. For the dissociation, the two-photon detuning is tuned to be
$220-320$ kHz above the bound-free transition, and the powers of both beams are
reduced to obtain half the peak intensity as was used in the association process. This
Raman dissociation light reduces the lifetime of the AC molecules to  around
$80~\mu$s, as is shown in Fig. \ref{figs4}b, and thus the AC molecules can be
quickly eliminated. We checked that the Raman removing pulse has negligible
effects on the C atoms. The Raman removing pulse also affects the lifetime of
the AB molecules. However, the AB molecule still has a lifetime of about 2.7
ms under the Raman removing fields, as is shown in Fig. \ref{figs4}d, which is long
enough for the study of the reaction dynamics.

\begin{figure}[ptbh]\centering
\includegraphics[width=8cm]{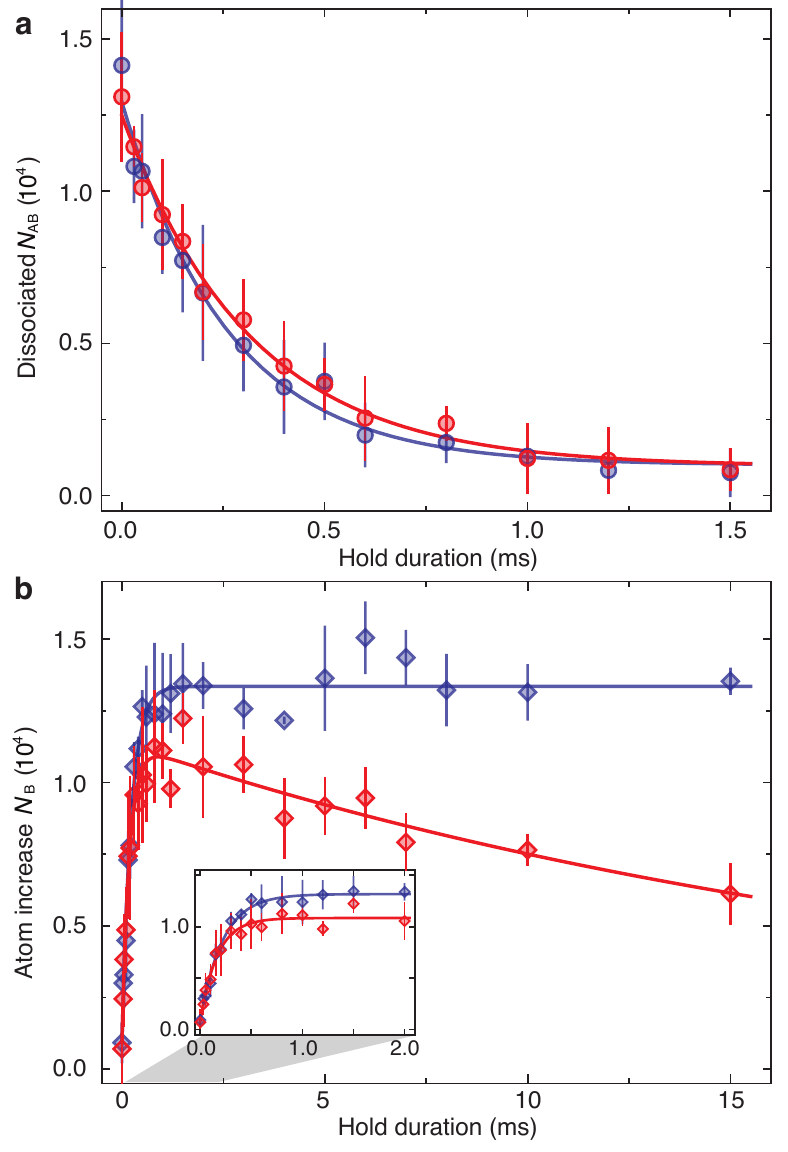} \renewcommand\thefigure{S5}
\caption{\textbf{Time evolution after removing the remaining A atoms.} The red (blue) data points represent the time evolution without (with) applying the Raman removing light at 130.26 G. \textbf{a}. The time evolution of the AB reactants. The time constants are measured to be $\tau_{\rm{AB}}=0.319(21)$ ms (red) and $\tau_{\rm{AB}}=0.262(28)$ ms (blue), respectively. \textbf{b}, The time evolutions of the B products, which are fitted with the model, $N_{\rm B}(t)=N^{0}_{\rm B}(e^{-t/\tau_2}-e^{-t/\tau_1})$. The fitting gives $\tau_1=0.184(24)$ ms and $\tau_2=21(4)$ ms (red), and $\tau_1=0.220(18)$ ms and $\tau_2=+\infty$ (blue).  \textbf{(Inset)}. The short term evolution of the B products.  The simple exponential fitting gives a time constant of $\tau_{\rm{B}}=0.166(22)$ ms (red), and $\tau_{\rm{B}}=0.220(18)$ ms (blue). Error bars represent $\pm1$ s.d. }\label{figs5}%
\end{figure}

We also compare the time evolutions when only removing the remaining A atoms
and the time evolutions when both the remaining A atoms and the AC molecules are removed. After removing the A
atoms, the dominant atoms in the mixture are the C atoms. The AC molecule has
a long lifetime of about 8 ms when coexisting with the C atoms, as shown in
Fig. \ref{figs4}a. In this case, the effects of the reverse reaction may also
be observed on a long time scale. As shown in Fig.\ref{figs5}, when only the A
atoms are removed, we observe a long-term decay for the B atom products.
However, this long-term decay disappears when the AC products are continuously
removed during the reaction process. Therefore, we attribute this long-term decay
to the influence of the reverse reaction.

\begin{figure}[ptbh]\centering
\includegraphics[width=8cm]{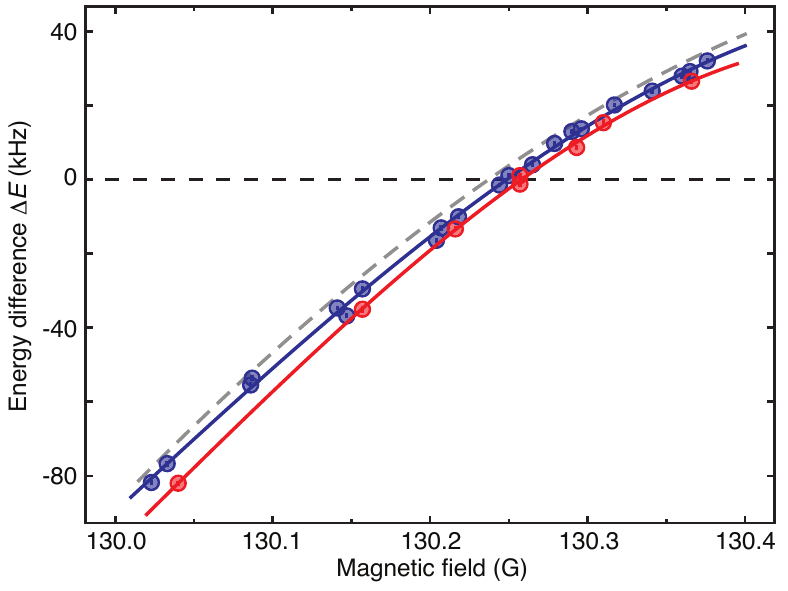} \renewcommand\thefigure{S6}
\caption{\textbf{Influence on the energy change of the Raman removing fields.} The red points represent the measured energy change under the Raman removing fields. The blue points represent the energy change in the main text. The solid lines are polynomial fitting to the measured values, which shows the zero-energy-change magnetic field is shifted by 9 mG under the Raman fields. The grey dashed line is the energy change calculated from the universal binding energy curves. Error bars inside the circles represent $\pm1$ s.d.}\label{figs6}%
\end{figure}

Besides, we also measure the influence of the Raman removing light on the
molecular and atomic transition frequencies. Our measurements show that the atomic
transition frequency $\nu_{\mathrm{C\rightarrow B}}$ is shifted by
about $+0.3$ kHz, whereas the molecular bound-bound transition frequency
$\nu_{\mathrm{AC\rightarrow AB}}$ is shifted by about $-3.4$ kHz. The energy
change $\Delta E=\nu_{\mathrm{AC\rightarrow AB}}-\nu_{\mathrm{C\rightarrow B}%
}$ under the Raman removing fields is plotted in Fig. \ref{figs6}. The
polynomial fitting shows the magnetic field of zero energy change is shifted
to 130.258(2) G under the Raman removing fields.

\subsection{Reaction dynamics}

Taking into account the reverse reaction, the time evolution of the AB
reactant and B product may be described by
\begin{align}
\dot{N}_{\mathrm{AB}}  &  =-\gamma_{\mathrm{AB}}N_{\mathrm{AB}}+k_{re}%
N_{\mathrm{AC}}N_{\mathrm{B}},\\
\dot{N}_{\mathrm{B}}  &  =\beta_{r}\bar{n}_{\mathrm{C}}N_{\mathrm{AB}}%
-k_{re}N_{\mathrm{AC}}N_{\mathrm{B}},
\end{align}
where the first and second term of the right side account for the
forward and reverse reaction respectively. The time evolution of the AC
molecule may be written in a similar form as the AB molecule. Since the time
evolution of the AC molecule is difficult to detect, its equation is not
shown. Here the overall loss rate of $N_{\mathrm{AB}}$ is $\gamma
_{\mathrm{AB}}=\beta_{\mathrm{A}}\bar{n}_{\mathrm{A}}+\beta_{\mathrm{C}}%
\bar{n}_{\mathrm{C}}+\beta_{r}\bar{n}_{\mathrm{C}}$, with $\bar{n}%
_{\mathrm{A}}$ and $\bar{n}_{\mathrm{C}}$ the mean densities of the A and C
atoms, respectively. The first term accounts for the loss due to inelastic
collisions with the remaining A atoms, with $\beta_{\mathrm{A}}$ the loss
rate coefficient. When the A atoms are removed, this term can be neglected. The other two
terms describe the losses due to collisions with the atom reactant C, which
include the desired reactive collision with a reaction rate coefficient of
$\beta_{r}$, and the losses due to reactions in other channels and vibrational
relaxations with an overall loss rate of $\beta_{\mathrm{C}}$. The collisions
between the AB molecules are neglected since the Feshbach molecules are
fermionic molecules. We have assumed $\bar{n}_{\mathrm{A}}$ and $\bar
{n}_{\mathrm{C}}$ are constants since the number of molecules is about one
order of magnitude smaller than that of the atoms. The reverse reaction is
caused by the reactive collisions between the AC molecule and B atom products
with $k_{re}$ the reverse reaction rate.

In our experiment, the time evolutions of $N_{\mathrm{AB}}(t)$ and
$N_{\mathrm{B}}(t)$ are measured. To extract the reaction rate coefficient, we use%

\begin{equation}
\beta_{r}=\frac{\dot{N}_{\mathrm{B}}(0)}{\bar{n}_{\mathrm{C}}N_{\mathrm{AB}%
}(0)}. \label{rate2}%
\end{equation}
This is reasonable as we start from the $\mathrm{AB+C}$ mixture and thus at
$t=0$ the number of products $N_{\mathrm{AC}}$ and $N_{\mathrm{B}}$ can be
neglected. The exact form of $N_{\mathrm{AB}}(t)$ and $N_{\mathrm{B}}(t)$ are
difficult to know. Therefore, we assume they may be approximated by
exponentials and use exponentials to fit the data points. The reaction rate
coefficients are extracted with $N_{\mathrm{AB}}(0)$ and $\dot{N}_{\mathrm{B}%
}(0)$ obtained from the fitting curves. Note that when the reverse reaction is
suppressed or can be neglected, the analytic solutions of $N_{\mathrm{AB}}(t)$
and $N_{\mathrm{B}}(t)$ are exponentials, and thus reaction rate coefficients
obtained from Eq. (\ref{rate2}) are exact.

The initial molecule number has to be corrected by taking into account the
finite rf dissociation rate \cite{Rui2017s}, since the reaction is fast. The
finite $\pi$-pulse transfer efficiency of the K atoms from the B to
$|9/2,-7/2\rangle$ states also needs to be included, which is about 90\% with
the Na atoms, and 95\% without the Na atoms. Finally, all uncertainties of the
parameters are included to give the uncertainty in the reaction rate coefficients.

\subsection{STM equations}

The scattering amplitudes satisfy the STM equations%
\begin{equation}
\begin{split}
t_{ii}(k,p,E)  &  =\frac{m_{\mathrm{{A}}}}{\pi\mu_{d}}\sqrt{\frac
{a_{\mathrm{{AC}}}}{a_{\mathrm{{AB}}}}}\int^{\Lambda}dq\frac{q}{2k}%
K(k,q,E) \\
              & \times D_{\mathrm{{AC}}}(q,E) t_{fi}(q,p,E)\\
t_{fi}(k,p,E)  &  =\frac{2\pi\hbar^{4}}{\mu_{d}^{2}\sqrt{a_{\mathrm{{AB}}%
}a_{\mathrm{{AC}}}}}\frac{m_{\mathrm{{A}}}}{2pk}K(k,p,E)\\
               &+\frac{m_{\mathrm{{A}%
}}}{\pi\mu_{d}}\sqrt{\frac{a_{\mathrm{{AB}}}}{a_{\mathrm{{AC}}}}}\int
^{\Lambda}dq\frac{q}{2k}K(k,q,E)\\
               & \times D_{\mathrm{{AB}}}(q,E)t_{ii}(q,p,E)
\end{split}
\end{equation}
where%
\begin{align*}
K(k,p,E)  &  =\log\frac{2\mu_{d}E-p^{2}-k^{2}+2\mu_{d}pk/m_{\mathrm{{A}}}%
}{2\mu_{d}E-p^{2}-k^{2}-2\mu_{d}pk/m_{\mathrm{{A}}}}\\
D_{\mathrm{{jl}}}(q,E)  &  =[-\frac{\hbar}{a_{\mathrm{{jl}}}}+\sqrt{-2\mu
_{d}(E-\frac{q^{2}}{2\mu_{ad}})}]^{-1}.
\end{align*}

We denote the input channel $\mathrm{{AB+C}}$ by channel $i$ and the reaction
channel $\mathrm{{AC+B}}$ by channel $f$, and thus $t_{ii}(k,p,E)$ is the
elastic scattering amplitude, or in other words, the back-reflection amplitude
of scattering into the input channel, and $t_{fi}(k,p,E)$ is the reactive
scattering amplitude, with $p$ and $k$ being the relative momentum of the
incoming and outgoing atom-dimer pairs and $E$ the total energy, $\mu
_{d}=m_{\mathrm{{A}}}m_{\mathrm{{B}}}/(m_{\mathrm{{A}}}+m_{\mathrm{{B}}})$ is
the reduced mass, and $\mu_{ad}=m_{\mathrm{{A}}}(m_{\mathrm{{A}}%
}+m_{\mathrm{{B}}})/(2m_{\mathrm{{A}}}+m_{\mathrm{{B}}})$ is the reduced of
mass of atom and dimer. In the integral equations, the only input parameters
are the scattering lengths $a_{\mathrm{{AB}}}$ and $a_{\mathrm{{AC}}}$, which
have been given in Sec. A in the Supplementary Information, and the unknown
three-body parameter $\Lambda$, which will be determined by comparing the
reaction rates. Since the magnetic field window in the experiment is narrow,
we assume $\Lambda$ is a constant.

We numerically solve the STM integral equations to calculate the on-shell
reaction amplitudes $t_{fi}(k_{f},p_{i},E)$ with $E=\hbar^{2}k_{f}^{2}%
/2\mu_{d}-E_{\mathrm{AC}}^{\mathrm{{b}}}=\hbar^{2}p_{i}^{2}/2\mu
_{d}-E_{\mathrm{AB}}^{\mathrm{{b}}}$ using the method introduced in Ref.
\cite{Braaten2009s,Helfrich2010s}. The incident relative momentum is calculated
for $p_{\min}\leq p_{i}\leq p_{\max}$, where $p_{\min}=0$ for the exothermic
reactions and $p_{\min}=\sqrt{2\mu_{d}|\Delta E|}$ for the endothermic
reactions, and $p_{\max}=$ $\sqrt{2\mu_{d}E_{\mathrm{AB}}^{\mathrm{{b}}}}$,
i.e., we only consider the collisional kinetic energy smaller than the binding
energy of the AB molecule. This is a good approximation since in our
experiment we have $E_{\mathrm{AB}}^{\mathrm{{b}}}>4k_{\mathrm{B}}T$. The
total reaction cross section is given by $\sigma_{r}(p_{i})=(8\pi^{3}%
/h^{4})\mu_{ad}^{2}(k_{f}/p_{i})|t_{fi}(k_{f},p_{i},E)|^{2}.$ The reaction
rate coefficient is thus calculated by $\beta_{r}=\int v\sigma_{r}(v)f(v)dv$
where we have $v=p/\mu_{ad}$ and  $f(v)=4\pi v^{2}[\mu_{ad}/(2\pi
k_{\mathrm{{B}}}T)]^{3/2}\exp[-(\mu_{ad}v^{2})/(2k_{\mathrm{{B}}}T)]$ is the
Maxwell-Boltzmann distribution. In the calculation, the temperature is assumed
to be $700$ nK, and the magnetic field has been shifted $10$ mG so that the
overlapping magnetic field obtained from the universal binding energy curves
is consistent with the direct measurement.

The calculated rate coefficients are compared with the experiment for the
magnetic fields between $129.44$ G to $130.38$ G, where the data points
between $129.44$ G and $129.91$ G are obtained from our previous work. The
reaction rate at $130.03$ G is also measured in our previous work
\cite{Rui2017s}. However, due to the suppression of the bound-free
Franck-Condon coefficients, the number of the AB molecule that can be formed
from the A+C mixture is only about 5000. In comparison, using the indirect
method in this work, $1.5\times10^{4}$ AB molecules can be formed. Therefore,
at 130.03 G, the reaction rate measured in this work has much better
signal-to-noise ratio and thus is used in the fitting. For the magnetic fields
larger than 130.03 G, the reaction rates can only be measured using the
current method.

\end{document}